# Time-resolved aortic 3D shape reconstruction from a limited number of cine 2D MRI slices

Gloria Wolkerstorfer[1], Stefano Buoso[1], Rabea Schlenker[1,2], Jochen von Spiczak[1,3], Robert Manka[1,2,3], Sebastian Kozerke[1]

*[1]Institute for Biomedical Engineering, University Zurich and ETH Zurich, Zurich, Switzerland*

*[2]Department of Cardiology, University Heart Center, University Hospital Zurich, University of Zurich, Zurich, Switzerland*

*[3]Diagnostic and Interventional Radiology, University Hospital Zurich, University of Zurich, Zurich, Switzerland*

Abstract

**Background and Objective:** To assess the feasibility and accuracy of reconstructing time-resolved, three-dimensional, subject-specific aortic geometries from a limited number of standard cine 2D magnetic resonance imaging (MRI) acquisitions. This is achieved by coupling a statistical shape model with a differentiable volumetric mesh optimization algorithm.

**Methods:** Cine 2D MRI slices were manually segmented and used to reconstruct subject-specific aortic geometries via a differentiable mesh optimization algorithm, constrained by a statistical shape model. Optimal slice positioning was first evaluated on synthetic data, followed by in-vivo acquisition in 30 subjects (19 volunteers and 11 aortic stenosis patients). Time-resolved aortic geometries were reconstructed, from which geometric descriptors and radial strain were derived. In a subset of 10 subjects, 4D flow MRI data was acquired to provide volumetric reference for peak-systolic shape comparison.

**Results:** Accurate reconstruction was achieved using as few as six cine 2D MRI slices. Agreement with 4D flow MRI reference data yielded a Dice score of (89.9 ± 1.6) %, Intersection over Union of (81.7 ± 2.7) %, Hausdorff distance of (7.3 ± 3.3) mm, and Chamfer distance of (3.7 ± 0.6) mm. The mean absolute radius error along the aortic arch was (0.8 ± 0.6) mm. Secondary analysis demonstrated significant differences in geometric features and radial strain across age groups, with strain decreasing progressively with age at values of (11.00 ± 3.11) × $10^{-2}$ vs. (3.74 ± 1.25) × $10^{-2}$ vs. (2.89 ± 0.87) × $10^{-2}$ for the young, mid-age, and elderly groups, respectively.

**Conclusion:** The proposed framework enables reconstruction of time-resolved, subject-specific aortic geometries from a limited number of standard cine 2D MRI acquisitions, providing a practical basis for downstream computational analysis.

Highlights:

- Generation of time-resolved 3D aortic computational meshes from a limited set of standard cine 2D MRI slices using differentiable statistical shape modelling and mesh optimization.
- Aortic shapes derived from as few as six standard cine 2D MRI slices compare well with shapes extracted from non-standard 4D flow MRI data.
- Quantitative metrics are readily derived including vessel shape parameters and radial wall strain.

**Graphical Abstract**

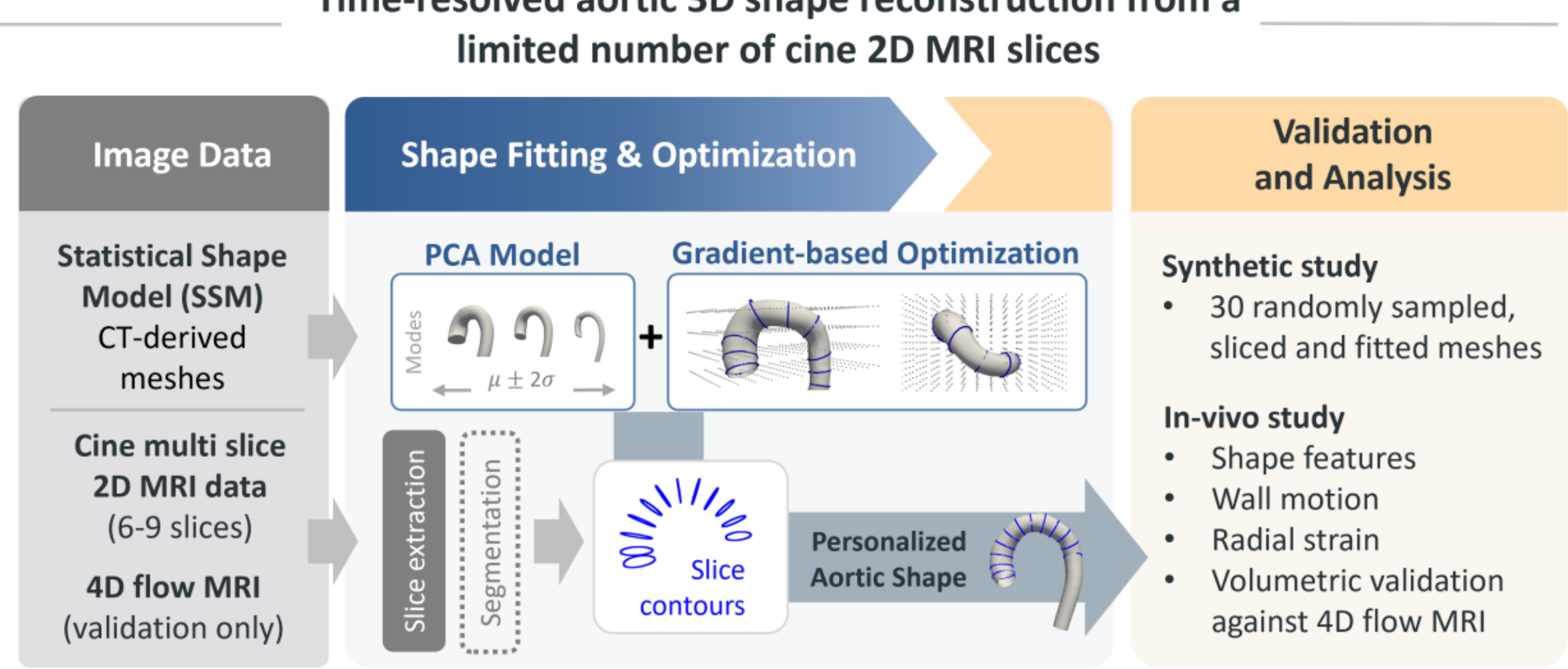

# 1 Introduction

Subject-specific computational models of cardiovascular anatomy are increasingly used for geometric analysis, biomechanical simulation, and individualized research applications [1–3]. In the context of the aorta, accurate reconstructions of time-resolved three-dimensional geometries are essential to study morphological variation [4–7], wall motion [8–10], disease-associated remodelling [11], and subject-specific in silico modelling [12]. Efficient methods for deriving dynamic 3D aortic geometries are therefore of growing interest [13, 14].

Statistical shape models (SSMs) capture anatomical variability through principal modes of shape variation. In the aorta, SSMs have primarily been developed from static volumetric Computed Tomography (CT) or MRI, to characterize morphology and pathology [15, 16]. However, subject-specific, time-resolved 3D reconstruction of the aorta from standard cine 2D MRI has not yet been established, despite related applications of SSM-based reconstructions in cardiac modelling [17, 18].

Multi-slice cine 2D balanced steady-state free precession (bSSFP) and gradient-echo (GRE) imaging are routinely used in clinical cardiovascular MRI to assess cardiac and vascular function and anatomy [19]. In contrast, volumetric time-resolved aortic imaging is commonly obtained only in research settings, most notably with 4D flow MRI [20, 21]. Although 4D flow MRI offers combined flow and shape assessment, its limited spatial (~2.5 mm$^3$) temporal and (40-50 ms) resolution, prolonged acquisition and reconstruction times, and reduced vessel-to-background contrast can constrain vessel wall delineation and motion analysis [22, 23]. Existing 4D flow MRI segmentation methods therefore often focus on peak-systole with reported Dice scores in the range of 0.83 to 0.92 on non-contrast data [24–29]. In addition, time-resolved 3D bSSFP approaches have also been explored [30], but are not yet part of routine clinical workflows.

The present work presents a framework for reconstructing subject-specific, time-resolved aortic geometries from a limited number of cine 2D MRI slices by coupling a statistical shape model to a differentiable mesh optimization. By leveraging routinely acquired sequences, the approach enables dynamic 3D reconstruction without requiring dedicated volumetric imaging. The primary objective of this study is to assess the feasibility and accuracy of geometric reconstruction. Secondary analyses, including radial strain estimation and shape characterization across age groups, are included to illustrate the potential utility of the method, but are neither the primary focus of the validation nor intended for clinical reference.

# 2 Materials and Methods

In this work, time-resolved cross-sectional segmentations derived from cine 2D MRI serve as input to a differentiable volumetric mesh optimization framework for generating subject-specific aortic geometries (Figure 1). The overall pipeline consists of three components: (i) construction of a statistical shape model from CT-derived data (A-B), (ii) acquisition and segmentation of cine 2D MRI slices, and (iii) subject-specific reconstruction via differentiable mesh optimization (D). Reconstruction performance was evaluated using synthetic data to assess behaviour under controlled conditions, and using in-vivo 4D flow MRI data for validation against volumetric reference data. Additionally, shape features and radial wall strains were extracted from the subject-specific MRI-based meshes and parameters compared for different age groups (E).

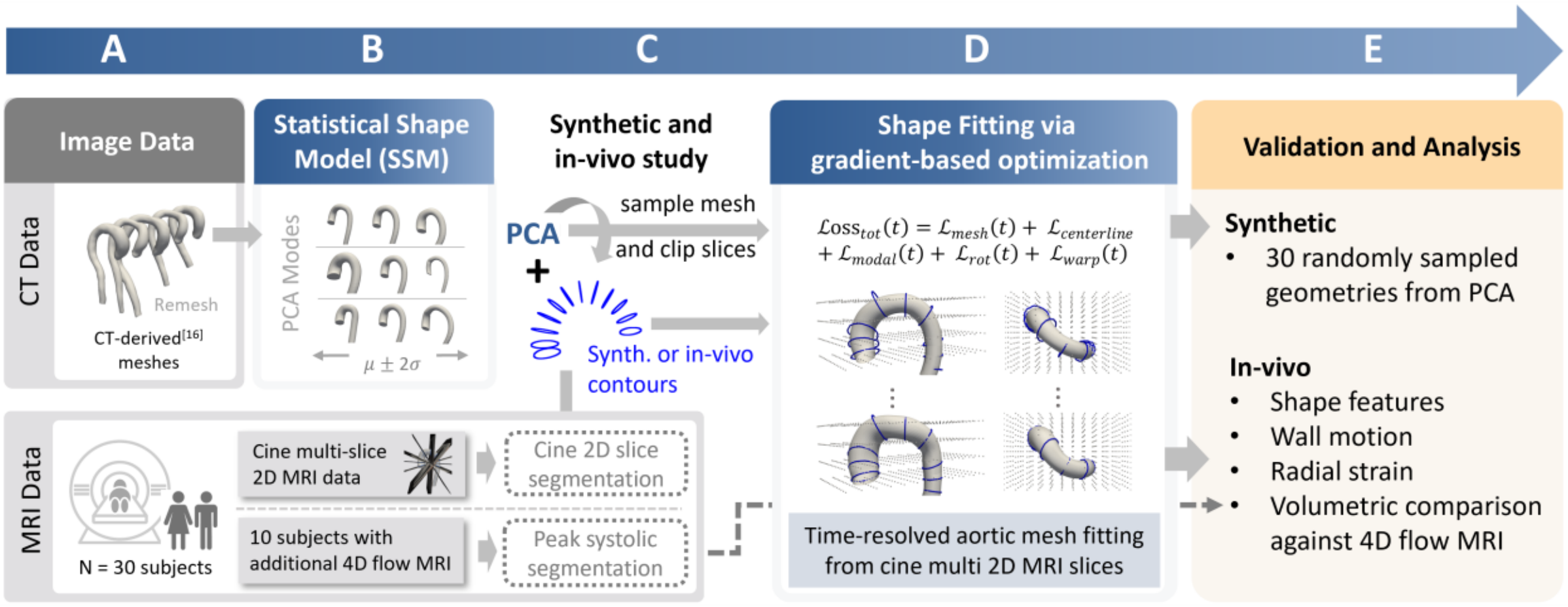

*Figure 1: Overview of the time-resolved aortic 3D shape reconstruction framework: (A-B) a statistical shape model and segmentations of standard cine multi-slice 2D MRI data acquired along the aorta serve as input (C) to a differentiable volumetric mesh optimization algorithm (D) yielding computational meshes suitable for various downstream tasks and shape analysis (E).*

## 2.1 Dataset for Statistical Shape Model

The statistical shape model was constructed from a publicly available dataset of 3000 synthetic aortic shapes derived from 26 CT-based meshes [16]. The dataset was chosen due to its high-quality mesh connectivity and anatomical coverage. To streamline processing and to exclude the aortic root, each mesh was clipped at the sinotubular junction and remeshed to a standardized grid with N = 3280 nodes and 6478 cells comprised in 40 cross-sections with 82 points each, ensuring a consistent topology. The SSM was generated using principal component analysis (PCA), capturing the primary modes of anatomical variation across the dataset. Weighted linear combinations of these PCA modes enabled geometric reconstruction. Figure 1B (top) illustrates the SSM's expressivity by demonstrating the effect of varying the weights of the three dominant modes. For the subsequent differentiable volumetric mesh optimization (Figure 1D), the first ten modes ($M = 10$), accounting for 98.5 % of the total variation, were included.

## 2.2 Shape fitting

Shape fitting was performed by adapting a differentiable mesh optimization framework originally proposed by Joyce et al. [17] for left ventricular modelling. The method couples a statistical shape model with volumetric mesh optimization to align the model to subject-specific 2D cross-sectional segmentations (Figure 1, C-D). The optimization proceeds by combining global shape variation captured by the SSM with local deformations driven by the observed 2D contours. The algorithm is outlined in Figure 2. In brief, the SSM mean shape, $\boldsymbol{X}_{mean} \in \mathbb{R}^{N\times 3}$, is first centered within a 3D Cartesian coordinate system. A total of $S$ segmented contours are extracted from the 2D image slices. Each contour is uniformly interpolated to $P = 180$ points, resulting in a combined set of slices $\boldsymbol{s} \in \mathbb{R}^{Q\times3}$, where $Q = S \times P$. These points are then positioned in the same 3D space (visualized as blue contours in Figure 1C). A regular grid of control points $\boldsymbol{c} \in \mathbb{R}^{K\times3}$, with $K = 720$, is initialized around the mesh (shown as gray points in Figure 1D). For each iteration $i$, the predicted mesh is updated by combining (i) global shape variations captured by the SSM and rigid transformations, as well as (ii) local deformations propagated from the surrounding control points. Specifically, for every time frame $t$ we first compute the mesh coordinates predicted by the SSM as:

$$\boldsymbol{X}_{PCA}(t) = \boldsymbol{X}_{mean} + \delta(t) \sum_{m=1}^{M} a_m(t)\, \sigma_{PCA_m} \boldsymbol{\Phi}_m,$$

where $\delta \in \mathbb{R}$ is a scalar, controlling deformation of the magnitude, and $\boldsymbol{a} \in \mathbb{R}^M$, $\boldsymbol{\sigma}_{PCA} \in \mathbb{R}^M$, $\boldsymbol{\Phi} \in \mathbb{R}^{M\times N\times3}$ are the amplitudes, standard deviations and basis vectors of the PCA model. The coordinates given by the SSM are then translated, rotated and scaled as:

$$\boldsymbol{X}_{TRANS}(t) = \psi(t)\, \boldsymbol{X}_{PCA}(t) \cdot \boldsymbol{R}(\alpha(t), \beta(t), \gamma(t)) + \boldsymbol{o}(t)\,,$$

where $\psi(t) \in \mathbb{R}$ determines an isotropic scaling, $\boldsymbol{R} \in \mathbb{R}^{3\times3}$ is the rotation matrix function of the Euler angles $(\alpha(t), \beta(t), \gamma(t))$, and $\boldsymbol{o}(t) \in \mathbb{R}^{N\times3}$ is a translation offset. Local deformations of point coordinates $\boldsymbol{x}(x, y, z) \in \boldsymbol{X}_{TRANS}$, are refined using radial basis function (RBF) interpolation with polyharmonic splines. The deformation field $RBF(\boldsymbol{x}, \boldsymbol{c}, \Delta\boldsymbol{c})$ can be modelled as:

$$RBF(\boldsymbol{x}, \boldsymbol{c}, \Delta\boldsymbol{c}) = \sum_{k=1}^{K} \omega_k \|\boldsymbol{x} - (\boldsymbol{c}_k + \Delta\boldsymbol{c}_k)\,\| + \mathbf{v}^T\tilde{\boldsymbol{x}}, \quad \text{where} \quad \tilde{\boldsymbol{x}} = \begin{bmatrix}\boldsymbol{x}\\1\end{bmatrix},$$

where $\boldsymbol{c}$ are the fixed control points, $\Delta\boldsymbol{c}$ their learned offsets, $\boldsymbol{\omega} \in \mathbb{R}^{K\times3}$ are the RBF weights, and $\mathbf{v} \in \mathbb{R}^{4\times3}$ is the linear coefficient matrix including bias. It should be noted that the weight and bias matrices are reinitialized and updated independently at every iteration.

The final mesh update $\boldsymbol{X}(t)$ is computed as:

$$\boldsymbol{X}(t) = \boldsymbol{X}_{TRANS}(t) + \mathrm{RBF}(\boldsymbol{X}_{TRANS}(t), \boldsymbol{c}, \Delta\boldsymbol{c}(t)).$$

Here, the only fixed parameters in the model are $\boldsymbol{X}_{mean}$, $\boldsymbol{\Phi}$, $\boldsymbol{\sigma}_{PCA}$ and the initial control point positions $\boldsymbol{c}$. All other parameters are optimized throughout the process. The optimization minimizes a composite loss function consisting of the data fidelity, centerline alignment and regularizations terms:

$$\mathcal{L}_{tot}(t) = \mathcal{L}_{mesh}(t) + \mathcal{L}_{centerline}(t) + \mathcal{L}_{modal}(t) + \mathcal{L}_{rot}(t) + \mathcal{L}_{warp}(t).$$

The mesh loss is defined by minimizing the squared $L_2$ norm distance between each of the mesh surface points $\boldsymbol{x}(t)$ and the closest 2D slice contours $\boldsymbol{s}(t)$:

$$\mathcal{L}_{mesh}(t) = \frac{1}{Q} \sum_{j=1}^{Q} \min_{x \,\in\, X(t)} \left\| \boldsymbol{s}(t)_j - \boldsymbol{x}(t) \right\|^2 .$$

The centerline loss, $\mathcal{L}_{centerline}$, enforces alignment between the mesh-derived centerline, $\boldsymbol{cl}_{mesh}(t) \in \mathbb{R}^{P_m\times3}$, interpolated to $P_m = 500$ points, and the data-derived centerline, $\boldsymbol{cl}_{slices}(t) \in \mathbb{R}^{P_s\times3}$, which is obtained by fitting a B-spline through the centers of mass of the segmented contours to $P_s = 300$ points (using SciPy [31]). It is defined as:

$$\mathcal{L}_{centerline}(t) = \frac{1}{P_s} \sum_{i=1}^{P_s} \min_{j \in P_m} \left\| \boldsymbol{cl}_{slices}(t)_i - \boldsymbol{cl}_{mesh}(t)_j \right\|^2 ,$$

implicitly optimizing parameters of scaling and translation. $\mathcal{L}_{modal}$, $\mathcal{L}_{rot}$ and $\mathcal{L}_{warp}$ are regularization losses minimizing the mean-squared modal amplitudes $\boldsymbol{a}(t)$, rotation angles $(\alpha(t), \beta(t), \gamma(t))$ and control point displacements $\Delta\boldsymbol{c}(t)$ as:

$$\mathcal{L}_{modal}(t) = \frac{1}{M}\sum_{m=1}^{M} a_m(t)^2 \,,$$

$$\mathcal{L}_{rot}(t) = \frac{1}{3}\sum \boldsymbol{R}(\alpha(t), \beta(t), \gamma(t))^2 \,,$$

$$\mathcal{L}_{warp}(t) = \frac{1}{K}\sum_{k=1}^{K} \Delta\boldsymbol{c}_k(t)^2 \,.$$

Due to the hierarchical significance of the different loss terms, we used selective activation of learnable parameters across a total of 300 epochs. Given spatiotemporal correlation over the cardiac cycle, shapes of subsequent cardiac frames were fitted using the last 50 epochs of the algorithm, initialized with each previously fitted one. The algorithm was implemented using PyTorch [32] and the Adam optimizer [33] with a learning rate of 0.1 was deployed.

**Algorithm 1:** Shape fitting pseudo code

**Input:** $\boldsymbol{X}_{\text{mean}}, \boldsymbol{\Phi}, \boldsymbol{a}^{\text{init}}, \boldsymbol{\sigma}_{\text{PCA}}, \delta^{\text{init}}; \boldsymbol{s}(t)$
**Output:** Personalized mesh $\boldsymbol{X}(t)$ and centerline $\boldsymbol{cl}_{mesh}(t)$
**for** *frame t in cardiac cycle* **do**
    Load cross-sectional slices $\boldsymbol{s}(t)$ and compute centerline $\boldsymbol{cl}_{slices}(t)$
    Center data by subtracting center of mass
    **if** $t = 0$ **then**
        Initialize `meshFitter` with SSM data and centered $\boldsymbol{c}$:
        $[\boldsymbol{a}(t_0^i), \delta(t_0^i)] \leftarrow [\boldsymbol{a}(t_0^{\text{init}}), \delta(t_0^{\text{init}})]$
        **for** *epoch i* **to** 300 **do**
            **if** $i < 10$: learn $\leftarrow [\boldsymbol{a}(t_0^i), \delta(t_0^i)]$
            **elif** 10 <i< 200: learn $\leftarrow [\boldsymbol{a}(t_0^i), \delta(t_0^i), \psi(t_0^i), \boldsymbol{R}(t_0^i), \boldsymbol{o}(t_0^i)]$
            **elif** 200 <i< 250: learn $\leftarrow [\boldsymbol{a}(t_0^i), \delta(t_0^i), \boldsymbol{R}(t_0^i), \Delta\boldsymbol{c}(t_0^i)]$
            **else**: learn $\leftarrow [\Delta\boldsymbol{c}(t_0^i)]$
            $\boldsymbol{X}_{\text{PCA}}(t_0^i) = \boldsymbol{X}_{\text{mean}} + \delta(t_0^i)\, \Sigma_{m=1}^{M}\, a_m(t_0^i)\, \sigma_{\text{PCA}_m} \boldsymbol{\Phi}_{m,n}$
            $\boldsymbol{X}_{\text{TRANS}}(t_0^i) = \psi(t_0^i) \cdot \boldsymbol{X}_{\text{PCA}}(t_0^i) \cdot \boldsymbol{R}(t_0^i) + \boldsymbol{o}(t_0^i)$
            $\boldsymbol{X}(t_0^i) = \texttt{RBF}(\boldsymbol{X}_{\text{TRANS}}(t_0^i), \boldsymbol{c}, \Delta\boldsymbol{c}_i^0)$
    **else**
        Fine-tune based on previous frame:
        **for** *epoch i* **to** 50 **do**
            learn $\leftarrow [\Delta\mathbf{c}(t_t^i)]$
            $\mathbf{X}(t_t^i) = \texttt{RBF}(\boldsymbol{X}(\mathrm{t}_{-1}^i), \boldsymbol{c}, \Delta\boldsymbol{c}(t_{-1}^i))$
    Predict: mesh points $\leftarrow$ model()
    Compute $\mathcal{L}_{\text{tot}}$ and update model

*Figure 2: Pseudo code for the shape fitting process.*

The SSM mesh processing and shape fitting code is available in the accompanying online repository.

## 2.3 In-vivo MRI data acquisition

Prospective MRI data was acquired in a total of 30 subjects across two sites using Philips MRI systems (Philips Healthcare, Best, the Netherlands). All subjects provided written informed consent, and data collection complied with institutional and ethical guidelines. Table 1 summarizes the description of the cohort population and sites. The subjects were categorized into three age groups: 11 young subjects (29 ± 4 years), 7 mid-aged (mid) subjects (63 ± 9 years), and 12 elderly subjects (82 ± 6 years). Eleven elderly subjects were previously diagnosed with aortic stenosis (AS).

*Table 1: Age-group description across the cohort.*

| Group | Age [years] | Male / Female | Field strength | Pathology |
|---|---|---|---|---|
| Young | 29 ± 4 | 5 / 6 | 1.5T | Healthy |
| Mid-age | 63 ± 9 | 5 / 2 | 1.5T (3); 3T (4) | Healthy |
| Elderly | 82 ± 6 | 9 / 3 | 3T | Aortic stenosis (11), Healthy (1) |

For each subject six to nine cine 2D MRI slices using either bSSFP or GRE were acquired approximately perpendicular to the aortic arch, using the MRI survey scan for planning. Image acquisition parameters were: 1.17-1.5 $mm^2$ in-plane resolution, 5-8 mm slice thickness and 25 ms temporal resolution (reconstructed to 40 temporal frames). The cine 2D slices were obtained evenly distributed across the arch, starting from one diameter upstream the sinotubular junction to the beginning of the abdominal aorta, with acquisitions acquired during individual breath-holds. In a subgroup of 10 subjects (3 young, 1 mid, 6 elderly; mean age 62 ± 23 years; 2 female), additional 4D flow MRI was acquired with 2.5 $mm^3$ isotropic spatial and 50 ms temporal resolution. Data acquisition involved prospectively ECG gating with continuous acquisition, followed by retrospective re-binning to account for heart rate variability, using a pseudo-spiral Cartesian undersampling scheme (acceleration factor R = 4-6.7) and with repeated sampling of $k_y = k_z = 0$ profiles. At 3T, data were collected using a 13-point velocity encoding (venc) scheme with venc settings of 50/250 cm/s for healthy (N = 3) and 150/350 cm/s for aortic stenosed subjects (N = 4). At 1.5T, data were acquired with a 4-point scheme using a venc of 150 cm/s in all directions. All datasets were reconstructed with a locally-low-rank approach [34] and multi-venc acquisitions were combined using Bayesian unfolding [35] and retaining the expiratory state for analysis. Peak systolic vessel segmentations were performed manually and with multiple iterative refinements over several days using ITK-SNAP [36].

The study cohort used for SSM and in-vivo analysis was assembled from multiple datasets. A comprehensive summary of all data sources contributing to the study is outlined in Table A2 (Appendix).

## 2.4 Shape fitting evaluation

*2.4.1 Synthetic validation*

The 30 meshes used for synthetic evaluation were generated by sampling from the statistical shape model (σ = 1.58 around the mean shape) and are therefore not fully independent from the data used to construct the PCA. Accordingly, this analysis is designed to assess the internal consistency of the reconstruction framework, rather than its ability to generalize beyond the learned shape distribution. While this setup does not avoid circularity in a strict sense, it provides a controlled setting to evaluate reconstruction accuracy under known conditions.

To avoid trivial correspondence between sampled meshes and input data, each geometry was intersected with independently defined orthonormal cross-sections along its centerline. The centerline was extracted and interpolated using cubic B-splines (SciPy implementation [31]). Based on the centerline length, twelve orthogonal cross-sections were generated along the aortic arch at uniform intervals of 1.3-1.8 cm, approximating the mean vessel radius.

Because the aortic root was excluded from the original statistical shape model, the mesh inlet corresponded to the sinotubular junction (STJ). For numerical stability during mesh intersection, the first cross-section was positioned 2.5 cm (approximately one aortic diameter) distal to the mesh's inlet. Subsequent cross-sections were then placed at uniform radial intervals along the arch. The second intersection coincided with the pulmonary artery (PA) plane, which is commonly used in clinical 2D flow MRI. This PA level (grey points indicated in Figure 5F) was therefore defined as the reference slice, encompassing intersections of both the ascending and descending aorta.

Shape reconstruction accuracy was quantified using Dice score (DSC), Intersection over Union (IoU), Hausdorff distance (HD), Chamfer distance, and relative radius errors evaluated at each cross-section. Metric definitions are provided in the Appendix.
To assess the influence of slice number and positioning, an iterative slice-selection strategy was employed. Starting from the reference plane (PA level with slice positions 2 and 12), additional slice positions were progressively added, based on minimizing the average error across all metrics. The resulting order reflects the relative importance of each slice location for global shape reconstruction.

For sparse sampling scenarios, centerline estimation was adapted accordingly. When fewer than five slices were available, a surrogate centerline was approximated using a semicircular fit between the first and last slices. Finally, interpolation accuracy between reconstructed meshes to sparsely sampled contours was evaluated by computing relative radius errors at all cross-sections for increasing numbers of acquisition slices.

### *2.4.2 In-vivo validation*

In-vivo validation was performed using 4D flow MRI as a practical volumetric reference, rather than an absolute ground truth. This evaluation assesses reconstruction performance under realistic imaging conditions.

Reconstructed peak-systolic aortic geometries were compared against manually labelled 4D flow MRI data. For voxel-based comparison, fitted meshes were projected onto the 4D flow MRI grid to compute Dice score and Intersection over Union. For surface-based evaluation, smooth surfaces were generated from the 4D flow MRI segmentations to compute Hausdorff and Chamfer distances. In addition, relative radius errors were evaluated along the aortic centerline using uniformly spaced cross-sections, consistent with the in silico analysis. To account for inter-scan misalignment between cine 2D MRI and 4D flow MRI acquisitions, volumetric masks were visually inspected and, where necessary, corrected using minor rigid translations.

## 2.5 In-plane vessel segmentation

Initial coarse aortic segmentations were obtained from the standard axial MRI survey scan using a conventional U-Net [37]. The segmented volume was skeletonized (using Scikit-learn) and fitted using a B-spline to extract the vessel centerline. This centerline was subsequently intersected with the cine 2D MRI slices to localize the aorta in the 2D slices, enabling cropping to 64 × 64 in-plane regions-of-interest. The nnU-Net [38] was trained to segment the aorta in the cropped cine 2D MRI images. For network training we utilized 22 subjects (healthy and stenosed), for which multiple acquisitions were available, resulting in a total of 150 datasets. Segmentation performance was evaluated on an independent test set comprising eight healthy subjects (mean age: 48 ± 22 years; 3 female subjects), scanned on the same 3T system. To minimize inter-scanner variability given the limited dataset size, the test set was restricted to acquisitions from a single scanner. In addition, four elderly stenosed subjects were exchanged with younger healthy volunteers to enable a publicly shareable test dataset and assess segmentation performance on a broader age range. To evaluate breath-hold position misalignments, we evaluated overlap between three subsequent breath-holds for each test subject. Evaluation metrics included temporal Dice score, Intersection over Union, Hausdorff distance and Average surface distance (ASD).

## 2.6 Vessel metrics

Following shape reconstruction, geometric and functional metrics as described in [16] were derived to characterize aortic morphology and wall motion. Each shape was orthogonally sliced along its centerline in 7.5 mm intervals. As illustrated in Figure 5F, the aorta was anatomically divided into three regions: ascending (A-B), arch-top (B-T-C) and descending (C-D). Geometric parameters including arch height (h), arch width (w) and centerline lengths were extracted. Aortic tortuosity was defined as $\mathrm{T} = 1 - \frac{\mathrm{w}}{\text{L-AD}}$, where L-AD is the centerline length from the ascending to the descending aorta.

Wall motion was quantified as the norm of the distance of mesh cell centers, calculated as the distance $d(p_0, p_t)$ between different cardiac phases $p_t$ (with time index $t$) and the diastolic reference state $p_0$. Relative radial vessel wall strains were evaluated as $\varepsilon = \frac{\Delta \mathrm{r}}{\mathrm{r}_0}$ with $\Delta \mathrm{r} = |\, \mathrm{r}_t - r_0|$, where $r_t$ and $r_0$ represent vessel radii at systolic phase $t$ and diastolic phase $t_0$, respectively. Maximum average strain values along the ascending aorta and maximum centerline extensions within L-AD were computed.

The distributions of all shape, strain, and centerline parameters were examined for normality and equality of variances within each age group. Based on these assessments, pairwise group comparisons were performed using either Student's t-test when variances were similar or Welch's t-test when variances differed.

# 3 Results

## 3.1 Shape fitting evaluation

### *3.1.1 Synthetic data*

For each combination of slices placed between positions 2 and 12 (see inset in Figure 3), we computed the Dice score, Intersection over Union, Hausdorff distance and Chamfer distance. Figure 3 illustrates how the number and positioning of slices affect the accuracy of the shape fitting process. The horizontal axis shows all candidate positions, sorted by the next optimal slice, while the vertical axis indicates the number of in silico "acquired" slices. At each iteration, the mean of all metrics was calculated, and the best-performing case (highlighted along the diagonal) was selected for the next iteration. Notably, slice position 1 was only considered after the first iteration. A mean ± standard deviation in Dice score of (94.8 ± 1.1) %, Intersection over Union of (90.1 ± 2.0 ) %, Hausdorff distance of (5.9 ± 2.3) mm and Chamfer distance of (2.5 ± 0.4) mm was achieved when including the 6$^{th}$ slice, as highlighted in red. Iteratively applying this selection strategy resulted in a final Dice score of (96.7 ± 0.3) %, Intersection over Union of (93.5 ± 0.5) %, Hausdorff- and Chamfer distances of (4.5 ± 1.7) mm and (2.0 ± 0.2) mm, respectively.

Radius error calculations were performed across the twelve cross-sections along the centerline for an increasing number of slices included in the shape reconstruction, referenced against the original shape. Results show a progressive reduction in relative radius errors with each iteration, see Figure 4, with only minor improvements from 7 slices upwards. Specifically, the total mean ($\mu$) and standard deviation ($\sigma$) of relative errors for six and eleven slices were (0.2 ± 1.1) × $10^{-1}$ and (-0.2 ± 1.2) × $10^{-3}$, respectively. Corresponding mean radius errors were found as (0.32 ± 1.70) mm for six acquisition slices and (0.03 ± 0.17) mm for eleven acquisition slices.

### *3.1.2 In-vivo data*

Evaluation against manually segmented 4D flow MRI data demonstrated agreement between the reconstructed and reference shapes. A mean Dice score of (89.9 ± 1.6) %, Intersection over Union of (81.7 ± 2.7) %, Hausdorff distance of (7.3 ± 3.3) mm and Chamfer distance of (3.7 ± 0.6) mm were calculated. Additionally, the analysis of relative radius errors based on uniformly spaced cross-sectional slices along the centerline resulted in an absolute radius error of (0.8 ± 0.6) mm, and a relative error of (3 ± 7) × $10^{-2}$, demonstrating acceptable shape fidelity throughout.

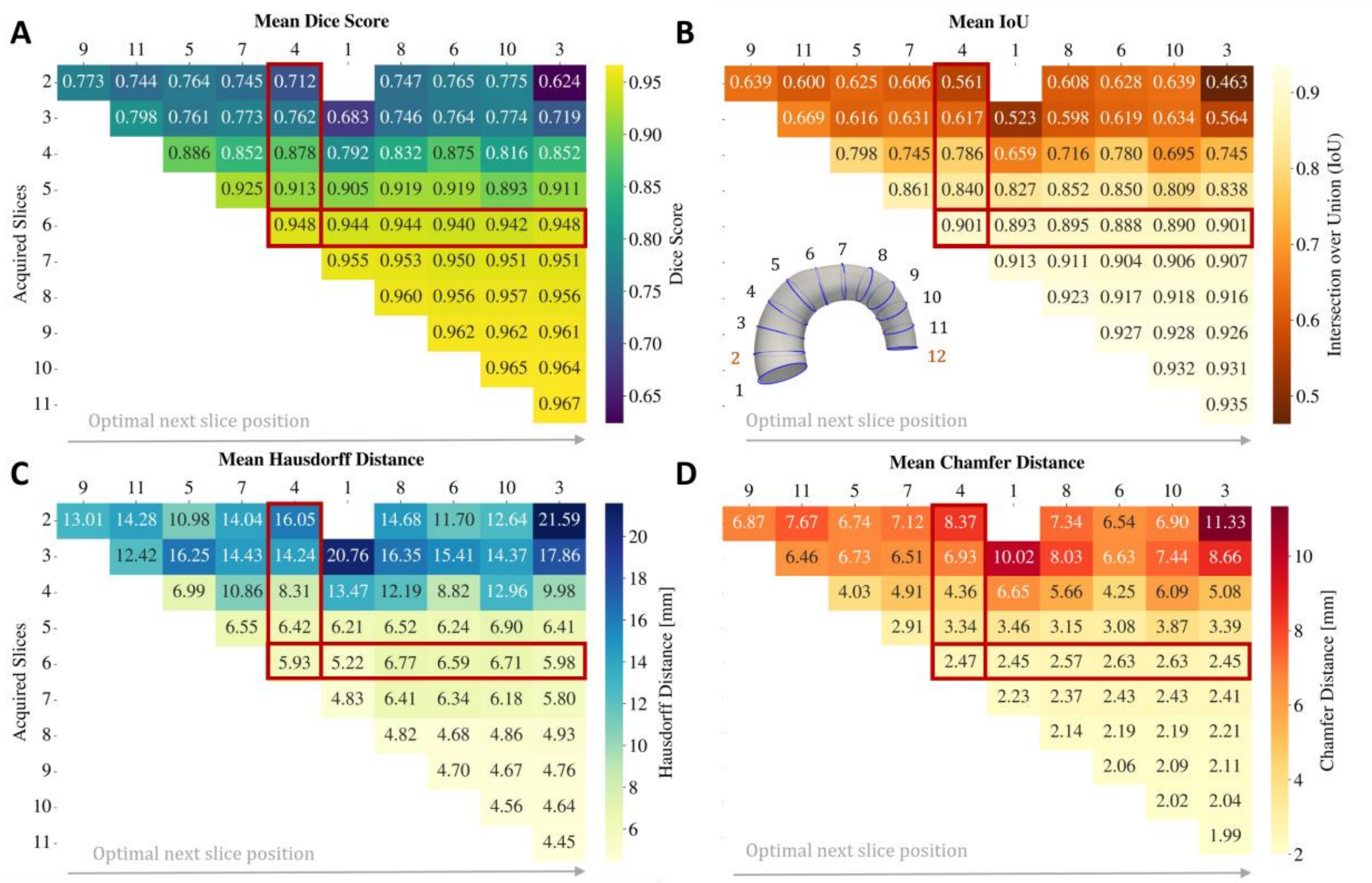

*Figure 3: Impact of slice number and positioning on mean Dice scores (A) and Intersection over Union (B), Hausdorff distance (C) and Chamfer distance (D) with the number of acquired slices and optimal next slice positions denoted along the vertical and horizontal axes, respectively. The slices yielding the highest mean metric were selected for the following iteration (diagonal). Red boxes highlight the desired Dice score of 95 %, with IoU 90 %, average Hausdorff distance at 6 mm and average Chamfer distance at 2.5 mm, when including 6 slices.*

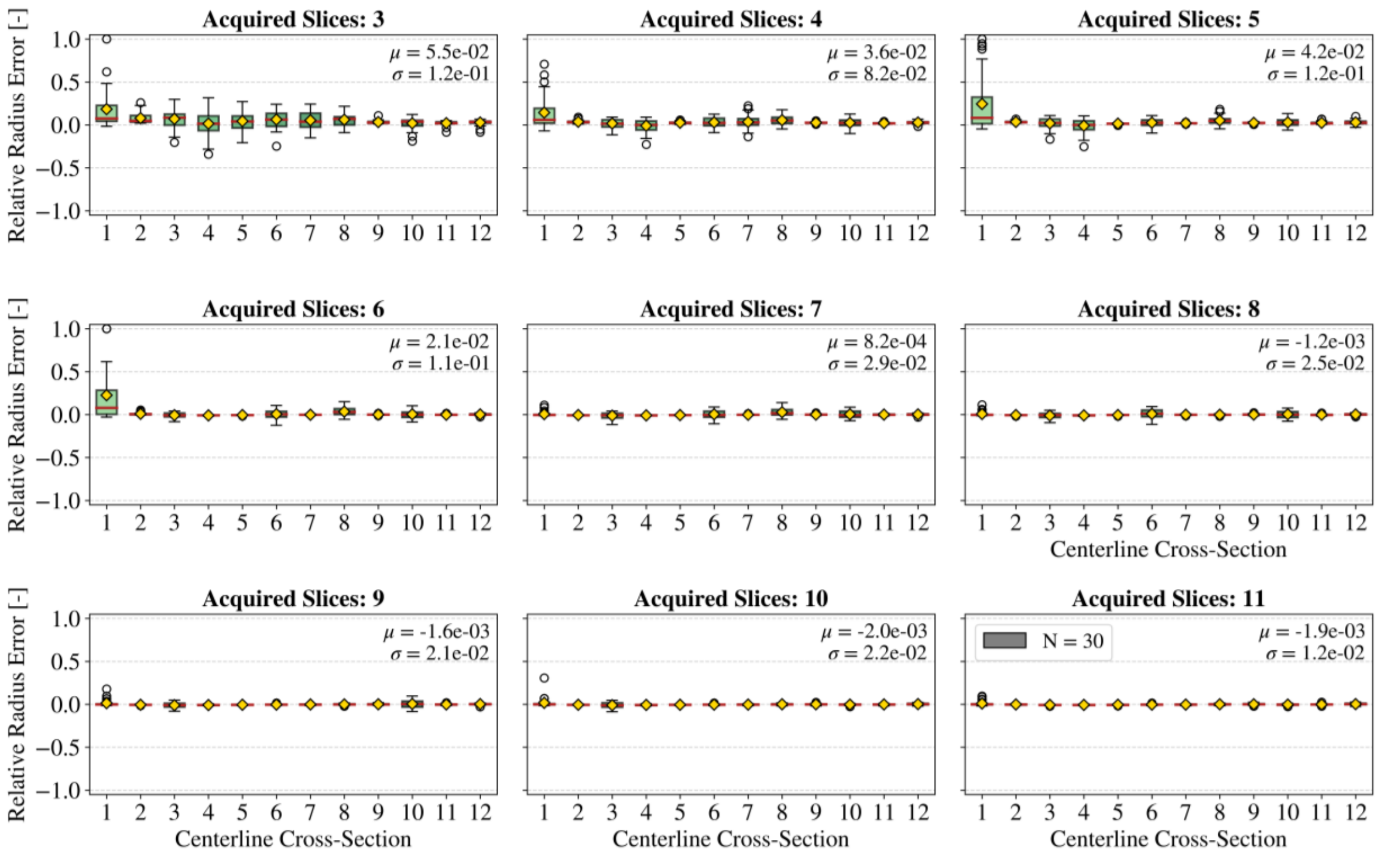

*Figure 4: Relative radius errors for increasing number of acquired slices. The relative radius errors of cross-sectional slices along the centerline are plotted for an increasing number of "acquired" slices. Resulting total mean relative errors for six and eleven slices were found as (0.2 ± 1.1) × $10^{-1}$ and (-0.2 ± 1.2) × $10^{-3}$, respectively.*

## 3.2 In-plane vessel segmentation

The nnU-Net model achieved high segmentation accuracy for both, the ascending and the descending aorta (see Appendix Figure A1). Table 2 summarizes the quantitative comparison of predicted segmentations against reference masks, reporting temporal averaged Dice scores, Intersection over Union, Hausdorff distance and average surface distance for ascending (AAo) and descending (DA) aorta, as well as the evaluation of averaged metrics from the three individual breath-holds as predicted by the network, referred to as MultiB-AAo and MultiB-DA.

*Table 2: Global distance metrics on the test dataset for ascending (AAo) and descending (DA) aorta with comparison between multiple breath-holds (MultiB) based on the network predictions.*

| Data | DSC ↑ | IoU ↑ | HD [mm] ↓ | ASD [mm] ↓ |
|---|---|---|---|---|
| **AAo** | 0.965 ± 0.012 | 0.933 ± 0.022 | 2.01 ± 0.30 | 0.20 ± 0.08 |
| **MultiB-AAo** | 0.957 ± 0.024 | 0.918 ± 0.042 | 2.03 ± 0.71 | 0.24 ± 0.11 |
| **DA** | 0.944 ± 0.023 | 0.894 ± 0.040 | 1.87 ± 0.36 | 0.26 ± 0.11 |
| **MultiB-DA** | 0.950 ± 0.025 | 0.907 ± 0.045 | 2.13 ± 2.14 | 0.22 ± 0.10 |

## 3.3 Vessel metrics

### *3.3.1 Shape feature analysis*

The comparison of shape features across age groups (young vs. mid-age vs. elderly) is presented in Figure 5. An increase in aortic size with age was measured in radii parameters for ascending (A): (11.1 ± 1.2) mm vs. (15.3 ± 0.9) mm vs. (16.6 ± 1.1) mm), top (B): (9.3 ± 1.2) mm vs. (12.2 ± 0.8) mm vs. (13.7 ± 1.1) mm, and descending (C): (8.4 ± 1.0) mm vs. (11.2 ± 0.8) mm vs. (12.5 ± 0.8) mm, respectively. In addition, ascending (D) and total (E) aortic arch lengths increased from (58.3 ± 9.6 and 120.9 ± 17.6) mm in the young vs. (77.8 ± 11.3 and 156.9 ± 15.2) mm in the intermediate vs. (97.3 ± 13.9 and 189.7 ± 27.7) mm in the elderly group. Aortic arch widths (G) and heights (H) also increased with age: (53.0 ± 5.7) mm vs. (73.4 ± 6.3) mm vs. (94.1 ± 11.2) mm and (45.8 ± 7.8) mm vs. (56.0 ± 6.9) mm vs. (65.2 ± 10.3) mm, respectively. Statistical testing revealed strong age-related differences ($p < 0.001$) between the young and elderly group for all parameters except tortuosity (I) ($p < 0.05$).

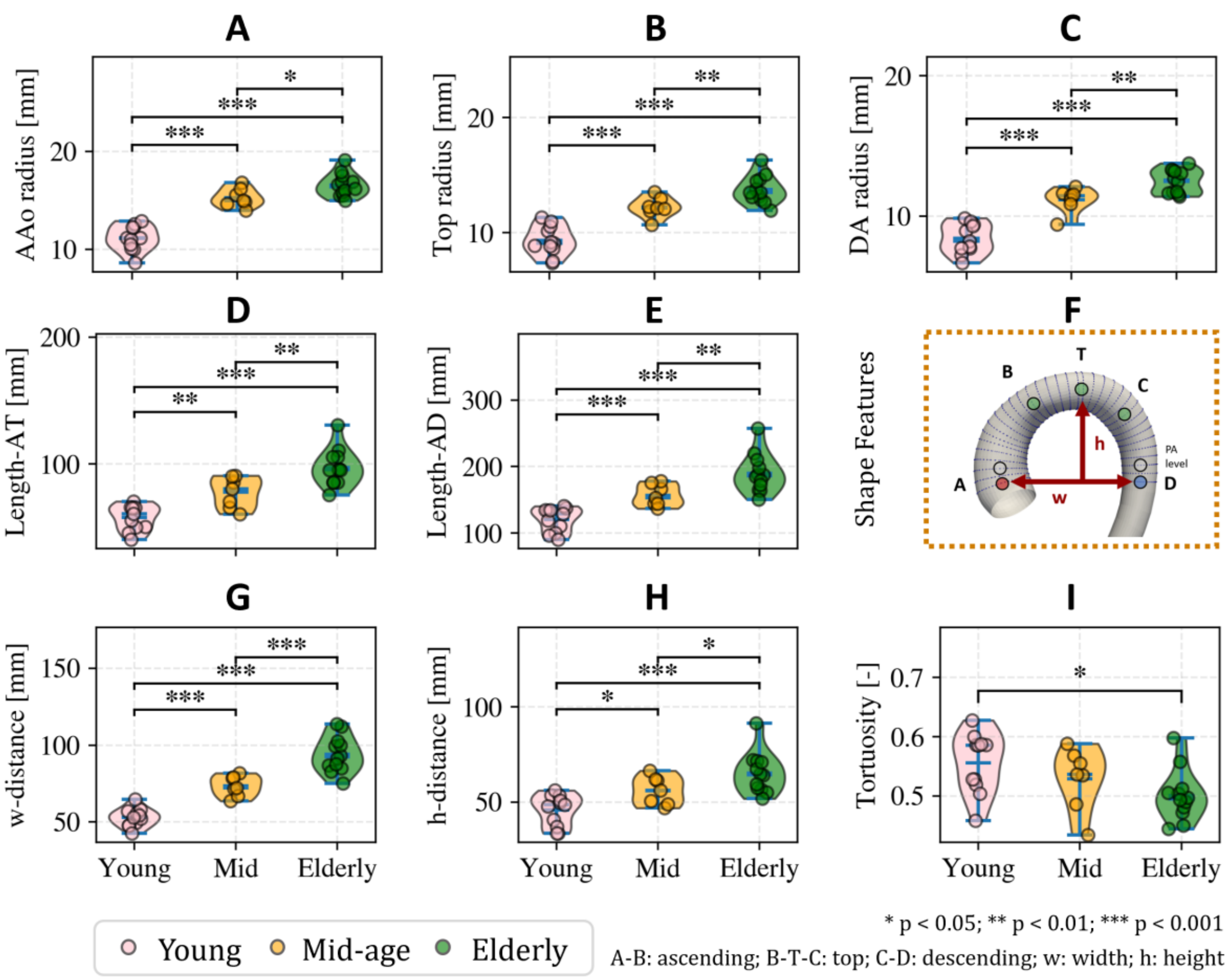

*Figure 5: Aortic shape feature comparison across age groups. Violin plots displaying the distribution of aortic shape features (F) with ascending aortic (AAo), top (Top) and descending aortic (DA) radii (A-C), ascending (D) and total (E) aortic arch length, width (G), height (H) and tortuosity (I) in young, mid-age, and elderly subjects. Significant increases in size were observed with age, with the largest differences between young and elderly groups ($p < 0.001$), except for tortuosity ($p < 0.05$).*

### *3.3.2 Wall deformation and strains*

Figure 6A illustrates relative wall motion across the three age groups (young, mid-age, elderly) in two examples for peak systole relative to end diastole. Peak systolic radial strains, averaged across the ascending aortic arch, are compared in Figure 6B. The data shows significant difference ($p < 0.001$) in radial strains between the young vs. mid-age and young vs. elderly groups, with mean and standard deviation values of (11.00 ± 3.11) × $10^{-2}$ vs. (3.74 ± 1.25) × $10^{-2}$ vs. (2.89 ± 0.87) × $10^{-2}$, respectively. Relative aortic centerline length change between diastole and peak systole were (3.5 ± 1.0) % vs. (1.9 ± 0.4) % vs. (1.4 ± 0.4) % with largest difference (p <

0.001) between young and elderly.

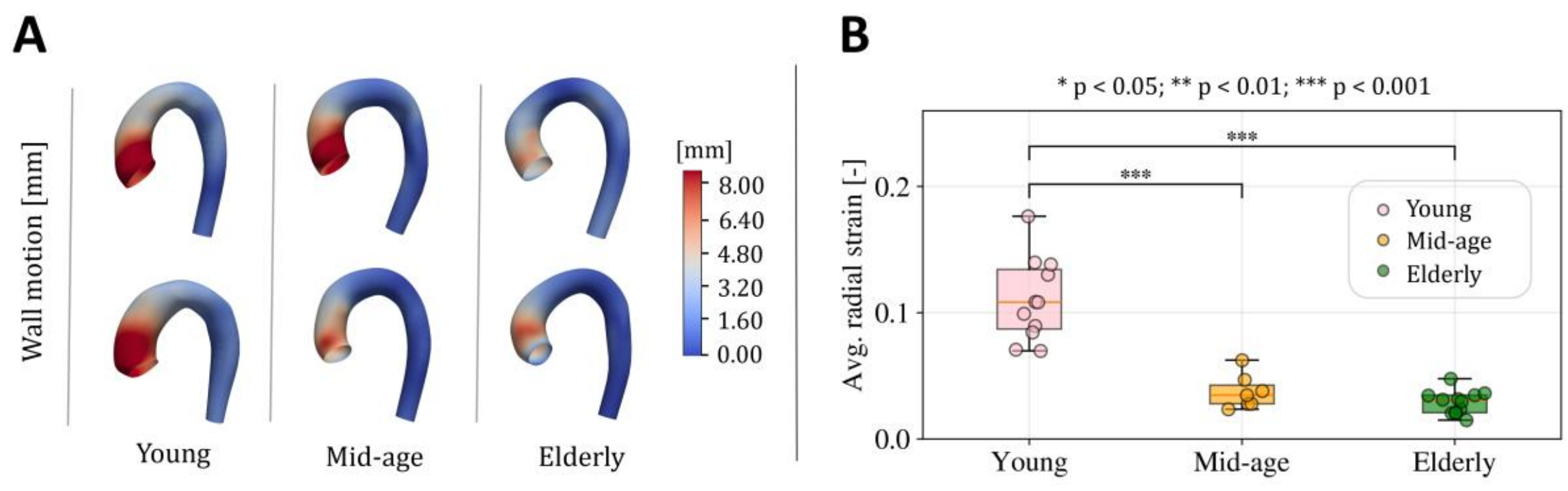

*Figure 6: Wall motion maps and radial strain analysis. (A) Wall motion across three age groups (young, mid-age, elderly) is illustrated with two examples each at peak systole. (B) Peak systolic radial wall strains, averaged across the ascending aortic arch.*

# 4 Discussion

A method for generating time-resolved, three-dimensional aortic meshes from a limited number of cine 2D MRI slices has been presented. Leveraging the high in-plane spatial and temporal resolution of standard cine MRI, accurate geometries were obtained from as few as six uniformly distributed slices along the arch.

Reconstruction was achieved using differentiable volumetric mesh optimization, combining an asymmetric Chamfer distance loss (referred to as "mesh loss") with a centerline constraint to ensure both geometric fidelity and stable convergence. Training parameters were optimized via grid search, to balance reconstruction accuracy and computational efficiency.

Synthetic analysis demonstrated that already six slices were sufficient to achieve high volumetric accuracy (Dice ~ 95 %), with mean radius errors below the effective CT resolution. In-vivo validation against 4D flow MRI showed good agreement, with Dice scores comparable to previously reported values (0.83 to 0.92) [26–29]. Chamfer distances and radius errors remained close to or below the image resolution throughout the aortic arch.

The observed Hausdorff distances (~4-6 mm in silico and ~7 mm in-vivo) should be interpreted in the context of the underlying imaging resolution. In-vivo errors are on the order of cine 2D MRI slice thickness (5-8 mm) and reflect partial-volume effects, segmentation uncertainty, and inter-scan misalignment, while exceeding the isotropic resolution of 4D flow MRI. Notably, the Hausdorff distance is sensitive to local outliers, therefore, complementary metrics such as the average surface distance provide a more robust assessment of overall shape agreement and remained close to the 4D flow MRI resolution.

While 4D flow MRI has inherent limitations in spatial resolution and vessel-to-background contrast, it was used here as a pragmatic volumetric reference for peak-systolic geometry comparison, rather than as an absolute ground truth. Slightly lower Dice agreement compared to [30] may be attributed to differences in spatial resolution and additional partial-volume effects during mesh-to-voxel sampling.

Although age- and disease-related effects cannot be separated in the present cohort, the observed trends are consistent with previously reported age-associated anatomical changes [5, 6, 16], demonstrating the framework's ability to reconstruct geometries across a heterogeneous age spectrum. Older subjects exhibited significantly larger aortic radii, longer centerlines, and increased arch heights and widths, as seen in literature. Time-resolved shape analysis enabled automatic estimation of wall motion, revealing reduced displacement, lower radial strain and relative centerline length change with increasing age, which is expected due to progressive aortic stiffening [7].

The nnU-Net-based segmentation demonstrated high temporal accuracy in both ascending and descending aortic cross-sections (Dice scores > 94 %), supporting the feasibility of automated, high-fidelity vessel segmentation. This is consistent with previous applications of the nnU-Net in volumetric aortic MRI [30, 39] and related approaches for cine 2D MRI segmentation [40, 41].

The proposed approach has several limitations. 4D flow MRI was used as a practical volumetric reference rather than an absolute ground truth, given its limited spatial resolution (~2.5 $mm^3$) and segmentation uncertainty. Accordingly, reconstruction errors should be interpreted relative to these imaging constraints. While computational phantoms (e.g. XCAT model [42, 43]) could provide access to ground-truth geometries, current implementations do not offer sufficiently detailed, time-resolved aortic models for this application, but represent a potential direction for future validation.

The statistical shape model was not evaluated on pathological anatomies such as aortic coarctations or dissections, limiting its immediate applicability to certain disease populations. Furthermore, while the combination of multiple 2D cine MRI slices provides a time-efficient alternative to volumetric imaging, it may be affected by patient motion, which can introduce shape uncertainty. Although slice positioning was consistent in this dataset, this cannot be assumed in routine clinical practice.

The current framework does not include the aortic root, despite its known relevance in valvular diseases. Extending the model to incorporate the root, either through direct modelling or integration of established root-specific shape models [44, 45], remains to be explored and validated, particularly in diseased populations [46].

Additionally, the elderly subgroup predominantly consisted of subjects with aortic stenosis, limiting the ability to disentangle age- and disease-related effects. Larger, age-, sex-, and pathology-stratified cohorts will be required to separate these contributions.

Finally, due to limited data availability, nn-UNet was used to assess the feasibility of automated segmentation rather than as part of the reconstruction pipeline, with all geometries derived from manually annotated contours. The network was trained and evaluated on a single scanner, and generalization to other vendors and field strengths remains to be established.

Overall, this work establishes the feasibility of the proposed framework and provides a foundation for scaling to larger, more diverse datasets, with potential applications in subject-specific biomechanical modelling and longitudinal analysis.

# 5 Conclusion

The proposed method enables the generation of time-resolved, subject-specific aortic geometries from a limited number of standard cine 2D MRI slices, allowing efficient extraction of dynamic 3D meshes suitable for shape and strain analyses, and downstream computational applications.

## Data availability

Our Python code for generating time-resolved subject-specific shape models from cine 2D MRI data is available on: https://gitlab.ethz.ch/ibt-cmr/publications/Temporal_Aortic_Shape_Twinning/.

### Declaration of competing interest

The authors declare that they have no known competing financial interests or personal relationships that could have appeared to influence the work reported in this paper.

## Acknowledgements

The authors acknowledge funding from the Swiss National Science Foundation (SNSF), grant CR23I3_166485, a Microsoft Joint Swiss Research grant and support from the Swiss Heart Foundation.

## Glossary (alphabetical)

- Ascending aorta (AAo)
- Average surface distance (ASD)
- Balanced-steady-state-free-precession (bSSFP)
- Computed Tomography (CT)
- Dice score (DSC)
- Descending aorta (DA)
- Gradient Echo (GRE)
- Hausdorff distance (HD)
- Intersection over Union (IoU)
- Magnetic Resonance Imaging (MRI)
- Principal Component Analysis (PCA)
- Radial basis function (RBF)
- Statistical shape model (SSM)

## CRediT author statement

**Gloria Wolkerstorfer:** Conceptualization, Methodology, Software, Validation, Formal analysis Investigation, Data Curation, Writing Original Draft, Writing Review Editing, Visualization; **Stefano Buoso:** Conceptualization, Methodology, Writing Original Draft, Writing Review Editing, Supervision; **Rabea Schlenker:** Resources, Writing Review Editing; **Jochen von Spiczak:** Writing Review Editing **Robert Manka:** Writing Review Editing**, Sebastian Kozerke:** Conceptualization, Methodology, Writing Original Draft, Writing Review Editing, Supervision, Project administration, Funding acquisition.

## AI use declaration:

During the preparation of this work the authors used ChatGPT for spellchecking and to improve the clarity of text. After using this tool, the authors reviewed and edited the content as needed and take full responsibility for the content of the publication.

# Appendix

### *A.1 In-plane vessel segmentation analysis*

Figure A1 illustrates temporal segmentation performance of the nnU-Net on the test subjects, with Dice scores of AAo and DA outlined in panel A. Predicted versus reference aortic radii at peak systole and one diastolic frame were compared using linear regression (panel B) and Bland-Altman analysis (panel C), which shows strong agreement between predicted and reference mask radii.

For the ascending aorta, the regression equation was $y = 0.93x + 0.90$ with r value of 0.98, a mean bias of -0.21 mm, confidence interval (CI) of [-0.45, 0.04] mm, and limits of agreement (LoA) of [-1.19, 0.78] mm. For the descending aorta, the results were $y = 0.90x + 0.9$, $r = 0.95$, with a mean bias of -0.46 mm, CI of [-0.75, -0.16] mm, and LoAs of [-1.64, 0.73] mm.

Repeatability analysis was conducted comparing segmentation predictions against reference segmentations from two annotators. Table A1 outlines global distance metrics between the initial reference annotation (R1 - as used in the manuscript) vs. second annotator (R2) vs. network predictions (P). Best segmentation agreement was found between the second annotator (R2) and network prediction, which we attribute to the expert's experience.

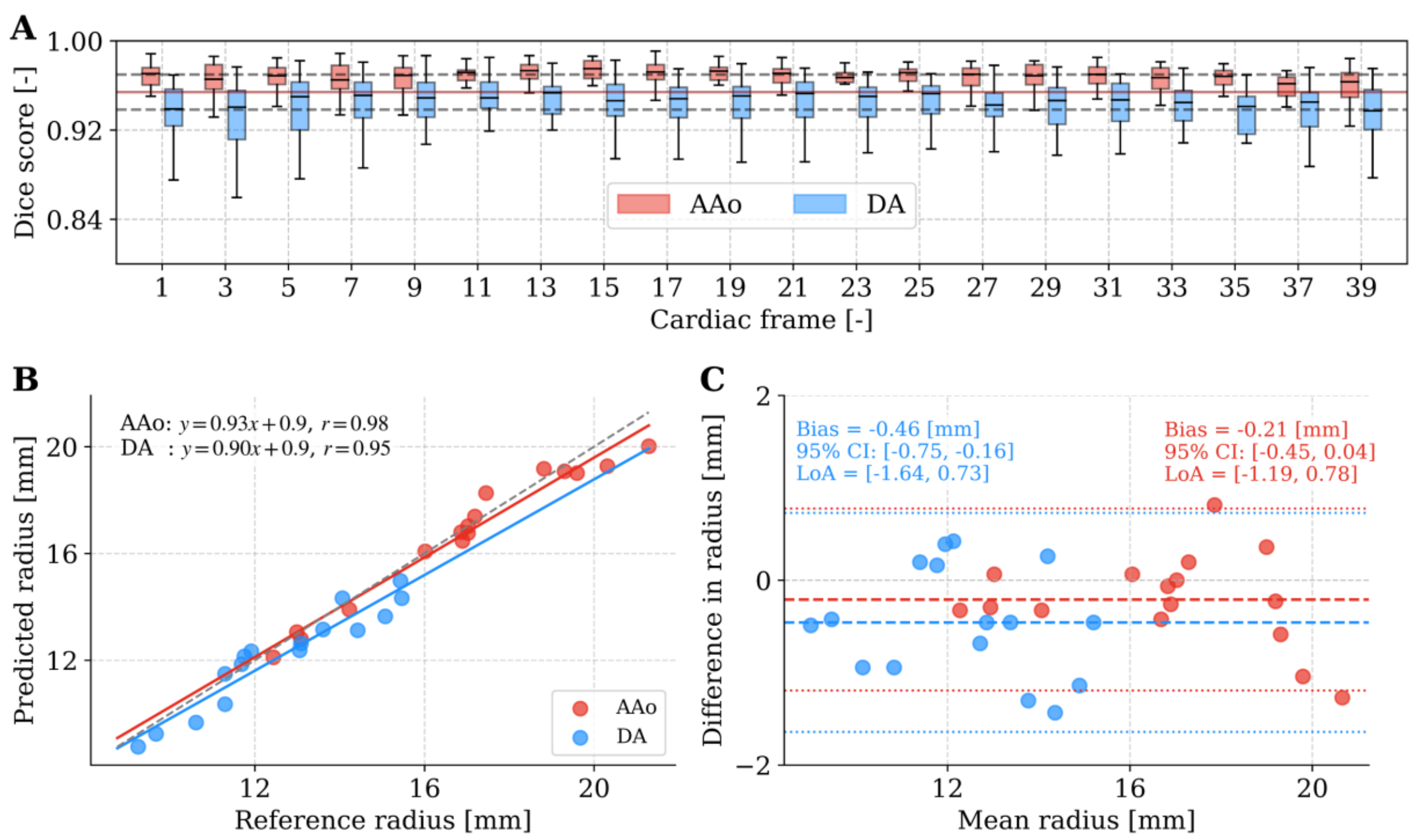

*Figure A1: Segmentation performance of the nnU-Net. (A) Temporal Dice scores in ascending (AAo in red) and descending (DA in blue) aorta across the cardiac cycle, for every interleaved frame. (B) Regression and (C) Bland-Altman analyses of radii at peak systolic and diastolic frame indicate strong agreement with manual annotations with displayed bias, confidence interval (CI), and limits of agreement (LoA) for AAo and DA, respectively.*

*Table A1: Global distance metrics between initial segmentation (R1) vs. second annotator (R2) and network predictions (P), for ascending (AAo) and descending aorta (DA), respectively.*

| Data | DSC ↑ | IoU ↑ | HD [mm] ↓ | ASD [mm] ↓ |
|---|---|---|---|---|
| **AAo** ($R_1$ vs. P) | 0.965 ± 0.012 | 0.933 ± 0.022 | 2.01 ± 0.30 | 0.20 ± 0.08 |
| **AAo** ($R_2$ vs. P) | **0.972 ± 0.003** | **0.946 ± 0.006** | **1.77 ± 0.36** | **0.16 ± 0.02** |
| **AAo** ($R_1$ vs. $R_2$) | 0.964 ± 0.007 | 0.930 ± 0.013 | 2.30 ± 0.29 | 0.20 ± 0.33 |
| **DA** ($R_1$ vs. P) | 0.944 ± 0.023 | 0.894 ± 0.040 | 1.87 ± 0.36 | 0.26 ± 0.11 |
| **DA** ($R_2$ vs. P) | **0.960 ± 0.011** | **0.923 ± 0.021** | **1.87 ± 0.62** | **0.19 ± 0.06** |
| **DA** ($R_1$ vs. $R_2$) | 0.948 ± 0.025 | 0.903 ± 0.043 | 1.84 ± 0.79 | 0.24 ± 0.12 |

### *A.2 Statistical Analysis*

To quantify the performance of binary segmentation masks, the number of true positives (TP), false positives (FP), true negatives (TN), and false negatives (FN) was used to compute the following voxel-wise metrics:

- Dice Score (DSC) quantifies the spatial overlap between predicted and reference segmentations:

$$\text{Dice} = \frac{2TP}{2TP + FP + FN}$$

ranging from 0 (no overlap) to 1 (perfect agreement).

- Intersection over Union (IoU) measures the ratio of the intersection to union:

$$\text{IoU} = \frac{TP}{TP + FP + FN}$$

also ranging 0 to 1, with stronger penalization for mismatches.

For surface-based evaluation, let $X$ and $Y$ denote the sets of boundary points extracted from the predicted and reference segmentations, respectively, with $d(x, y)$ defined as the Euclidean distance between points. Then the following point-set metrics can be defined:

- The Hausdorff Distance (HD) quantifies the maximum surface discrepancy:

$$\text{HD}(X, Y) = \max(h(X, Y), h(Y, X)),$$

where $h(A, B)$ is given by

$$h(A, B) = \max_{a \in A} \left( \min_{b \in B} (d(a, b)) \right).$$

- The Average Surface Distance (ASD) quantifies the mean bi-directional surface discrepancy as:

$$\text{ASD}(X, Y) = \frac{1}{|X| + |Y|} \left( \sum_{x \in X} \min_{y \in Y} d(x, y) + \sum_{y \in Y} \min_{x \in X} \left( d(x, y) \right) \right),$$

and is less sensitive to outliers than the HD.

- The Chamfer Distance (CD) quantifies the mean bi-directional surface discrepancy as:

$$\text{CD}(X, Y) = \frac{1}{|X|} \sum_{x \in X} \min_{y \in Y} d(x, y) + \frac{1}{|Y|} \sum_{y \in Y} \min_{x \in X} \left( d(x, y) \right),$$

which is similar to the ASD, but with different normalization.

*A.3 Data sources and cohort composition*

The study cohort was assembled from multiple datasets comprising CT-derived aortic meshes and in-vivo cine 2D MRI and 4D flow MRI data. To ensure transparency regarding data provenance and cohort composition, a summary of all contributing datasets is provided in Table A2.

*Table A2: Overview of datasets used for model development, validation and in-vivo analysis with indication of dataset source, sample size (N), role in analysis and degree of independence.*

| No. | Dataset | Source | N | Purpose | Independence |
|---|---|---|---|---|---|
| 1 | SSM training data | CT-derived meshes[16] | 3000 | Construction of statistical shape model (PCA) | Independent |
| 2 | Synthetic validation data | Sampled from SSM | 30 | Synthetic validation of reconstruction accuracy | Derived from SSM (not independent) |

| 3 | Cine 2D MRI data (in-vivo) | Internal (1.5T) | 14 | Input data for subject-specific shape reconstruction | Independent |
|---|---|---|---|---|---|
| 4 | Cine 2D MRI data (in-vivo) | Internal (3T) | 16 | Input data for subject-specific shape reconstruction | Independent |
| 5 | 4D flow MRI data (subset) | Internal (1.5T and 3T) | 10 | Volumetric reference for validation | Partially overlapping cohort with datasets 3-4. |
| 6 | Segmentation training data | Internal (1.5T and 3T) | 150 datasets (22 subjects) | Training of nnU-Net for automatic aortic segmentation | Fully overlapping cohort with datasets 3-4 |
| 7 | Segmentation test data | Internal (3T) | 8 subjects | Evaluation of segmentation performance | Partially overlapping cohort with datasets 3-4 |